\documentclass[pra,aps,epsf,nofootinbib,twocolumn,
notitlepage,showpacs,10pt, unsortedaddress, superscriptaddress]{revtex4-1}

\usepackage{color}
\usepackage{float}
\usepackage{graphicx}
\usepackage{subcaption}
\usepackage{amsmath}
\usepackage{amsthm}
\usepackage{mathtools}

\theoremstyle{definition}

\usepackage{bm}
\usepackage{hyperref}
\usepackage{float}
\usepackage{multirow}
\restylefloat{table}

\newcommand{\bra}[1]{{\left\langle #1 \right|}}
\newcommand{\ket}[1]{{\left| #1 \right\rangle}}

\theoremstyle{definition}
\newtheorem{thm}{Theorem}[section]

\usepackage{tikz} 
\usetikzlibrary{matrix,positioning}
\usepackage{amsfonts}
\usepackage{amssymb}
\newcommand{\rh}[1]{{\textcolor{blue}{#1}}} 

\hypersetup{
   colorlinks = true,
    urlcolor=blue
}
\usepackage[normalem]{ulem}
\usepackage{lipsum}

\newcommand\blfootnote[1]{%
  \begingroup
  \renewcommand\thefootnote{}\footnote{#1}%
  \addtocounter{footnote}{-1}%
  \endgroup
}

\begin{document}

\markboth{Rebekah Herrman}
{Multi-angle QAOA}

\title{Multi-angle Quantum Approximate Optimization Algorithm}

\author{Rebekah Herrman}
\email{rherrma2@tennessee.edu}\thanks{corresponding author, co-first author}
\affiliation{
	Department of Industrial and Systems Engineering\\ University of Tennessee at Knoxville\\Knoxville, TN 37996}
	
	\author{Phillip C. Lotshaw}
\email{lotshawpc@ornl.gov}\thanks{corresponding author, co-first author}
\affiliation{
	Quantum Computing Institute\\ Oak Ridge National Laboratory\\ Oak Ridge, TN 37830}
	
\author{James Ostrowski}
\affiliation{
	Department of Industrial and Systems Engineering\\ University of Tennessee at Knoxville\\Knoxville, TN 37996}

\author{Travis S. Humble}
\affiliation{
	Quantum Computing Institute\\ Oak Ridge National Laboratory\\ Oak Ridge, TN 37830}

\author{George Siopsis}
\affiliation{
	Department of Physics and Astronomy\\ University of Tennessee at Knoxville\\Knoxville, TN  37996}

\begin{abstract}
The quantum approximate optimization algorithm (QAOA) generates an approximate solution to combinatorial optimization problems using a variational ansatz circuit defined by parameterized layers of quantum evolution. In theory, the approximation improves with increasing ansatz depth but gate noise and circuit complexity undermine performance in practice. Here, we introduce a multi-angle ansatz for QAOA that reduces circuit depth and improves the approximation ratio by increasing the number of classical parameters. Even though the number of parameters increases, our results indicate that good parameters can be found in polynomial time. This new ansatz gives a 33\% increase in the approximation ratio for an infinite family of MaxCut instances over QAOA. The optimal performance is lower bounded by the conventional ansatz, and we present empirical results for graphs on eight vertices that one layer of the multi-angle anstaz is comparable to three layers of the traditional ansatz on MaxCut problems. Similarly, multi-angle QAOA yields a higher approximation ratio than QAOA at the same depth on a collection of MaxCut instances on fifty and one-hundred vertex graphs. Many of the optimized parameters are found to be zero, so their associated gates can be removed from the circuit, further decreasing the circuit depth. These results indicate that multi-angle QAOA requires shallower circuits to solve problems than QAOA, making it more viable for near-term intermediate-scale quantum devices. 


\end{abstract}

\maketitle

 \blfootnote{This manuscript has been authored by UT-Battelle, LLC under Contract No. DE-AC05-00OR22725 with the U.S. Department of Energy. The United States Government retains and the publisher, by accepting the article for publication, acknowledges that the United States Government retains a non-exclusive, paid-up, irrevocable, world-wide license to publish or reproduce the published form of this manuscript, or allow others to do so, for United States Government purposes. The Department of Energy will provide public access to these results of federally sponsored research in accordance with the DOE Public Access Plan. (http://energy.gov/downloads/doe-public-access-plan).}

\section{Introduction}
Among several quantum algorithms implemented on noisy intermediate-scale quantum (NISQ) devices \cite{cruz2019efficient, zhang2020efficient, godfrin2017operating, borle2020quantum, karamlou2020analyzing, hempel2018quantum, alderete2020quantum, xue2009quantum, linke2017experimental,pagano2020quantum, bengtsson2019quantum, harrigan2021quantum}, the quantum approximate optimization algorithm (QAOA) offers an opportunity to approximately solve combinatorial optimization problems such as MaxCut, Max Independent Set, and Max k-cover \cite{farhi2014quantum, saleem2020, cook2020quantum, crooks2018performance, pichler2018quantum, farhi2020quantumwholegraph, farhi2020quantum, wurtz2020bounds, ward2018qaoa, shaydulin2019evaluating}.   QAOA tunes a set of classical parameters to optimize the cost function expectation value for a quantum state prepared by well-defined sequence of operators acting on a known initial state. Variations to the original algorithm include alternative operators and initial states \cite{bartschi2020grover, hadfield2019quantum, wurtz2021classically, wang2019xy, egger2021warm, zhu2021improving, tate2020bridging, rieffel2020xy, golden2021threshold} while purely  classical aspects such as the parameter optimization and problem structure have been tested as well \cite{herrman2021impact, shaydulin2020classical, alexeev2020reinforcement, wauters2020reinforcement, lotshaw2021bfgs}. However, an outstanding concern is that practical implementations of QAOA require large numbers of qubits and deep circuits \cite{guerreschi2019qaoa,herrman2021lower}. Noise grows rapidly with circuit depth and affects the fidelity of the prepared quantum state \cite{xue2021effects, wang2020noise,murali2019noise, sun2021mitigating, marshall2020characterizing, alam2020design, alam2019analysis, streif2021quantum, maciejewski2021modeling}. 
\par
One approach to reduce the circuit depth of QAOA is to increase  
the number of classical parameters  introduced in each layer, a variation that we term multi-angle QAOA (ma-QAOA). Increasing the number of classical parameters allows for finer-grain control over the optimization of the cost function and the approximation ratio, which measures optimality relative to the known best solution. While introducing more classical parameters can lead to a more challenging optimization, a corresponding reduction in circuit depth preserves the critical resource of the quantum state. In addition, finding the absolute optimal angles is not necessary in order to see an improvement over QAOA. 

 Here, we quantify the advantages of using multiple parameters for each layer of QAOA. First, we prove that the approximation ratio converges to one as the number of iterations of ma-QAOA tends to infinity, a property that  ensures the optimal solution is the most likely. We next demonstrate that one iteration of ma-QAOA gives an approximation ratio that is at least that of the approximation ratio after one iteration of QAOA. This shows that ma-QAOA performs at least as well as QAOA. We also show that ma-QAOA used to solve the MaxCut problem on star graphs achieves an approximation ratio of one after one iteration, while single-iteration QAOA tends to an approximation ratio of 0.75 as the number of vertices goes to infinity. This result gives a concrete example where ma-QAOA gives a strictly larger approximation ratio than QAOA. We simulate solving MaxCut using ma-QAOA and QAOA on all connected, non-isomorphic eight vertex graphs and compare the performance of the two ansatzes. In doing so, we find that the average approximation ratio for ma-QAOA after one iteration is larger than the average approximation ratio of QAOA after three iterations. In looking at larger, 50- and 100-vertex graphs, we see that ma-QAOA retains its advantage over QAOA, giving approximation ratios that are on average six percentage points higher after the first iteration.

\section{Results} \label{results}

\subsection*{Quantum approximate optimization algorithm}\label{background}

The quantum approximate optimization algorithm (QAOA) relies on a combination of classical parameter optimization and applying cost and mixing operators to a quantum state in order to approximately solve combinatorial optimization (CO) problems \cite{farhi2014quantum}. The algorithm is implemented by applying two operators,
\begin{equation*}
U(C, \gamma) = e^{-iC\gamma}
\end{equation*}

\noindent and 
\begin{equation*}
U(B, \beta) = e^{-iB\beta}
\end{equation*}

\noindent in succession on an initial state, 
\begin{equation*}
\ket{s}=\frac{1}{\sqrt{2^n}} \sum_{z} \ket{z}.
\end{equation*}

\noindent Here the sum is over the computational basis $\ket{z}$. QAOA applied $p$ times to $\ket{s}$ is denoted $p$-QAOA. The state for $p$-QAOA is
\begin{equation*}
\ket{\gamma,\beta}=U(B, \beta_p)U(C,\gamma_p)...U(B, \beta_1)U(C,\gamma_1)\ket{s}.
\end{equation*}

\noindent $C$ encodes the problem to be solved and $B$ drives transitions between computational basis states. Often, $C$ is the sum over a collection of clauses, 

\begin{equation*}
C = \sum_{a} C_a,
\end{equation*}
\noindent and $B$ is typically 

\begin{equation*}
B = \sum_{v \in V(G)} B_v
\end{equation*}

\noindent where $B_v = \sigma_v^x$ is the Pauli-x operator acting on the $v^{th}$ qubit. Thus, we may write 

\begin{equation*}
U(C, \gamma) = e^{-i\gamma \sum_{a}C_a }
\end{equation*}

\noindent and 

\begin{equation*}
U(B, \beta) = e^{-i\beta \sum_{v \in V(G)}B_{v}}.
\end{equation*}

 Instead of focusing on minimizing the classical optimization efforts in QAOA, we modify QAOA such that it requires more classical parameters. The new classical parameters are introduced to QAOA by allowing each summand of the cost and mixing operators to have its own angle instead of a single angle for the cost operator and a second angle for the mixing operator. In this modification,
\begin{equation*}
U(C, \vec{\gamma_l}) =  e^{-i \sum_{a}C_a\gamma_{l,a} } = \prod_{a}e^{-i\gamma_{l,a} C_a }
\end{equation*}
\noindent and  
\begin{equation*}
U(B, \vec{\beta_l}) = e^{-i \sum_{v \in V(G)}B_{v}\beta_{l,v} } = \prod_{v \in V(G)}e^{-i\beta_{l,v} B_v }
\end{equation*}
\noindent where $\vec{\gamma_l} = (\gamma_{l,a_1}, \gamma_{l,a_2}, ... )$ and $\vec{\beta_l} = (\beta_{l,v_1}, \beta_{l,v_2}, ... )$. Here, $l$ denotes the layer, $a_i$ denotes an edges in the graph, and $v_j$ refers to a specific vertex. We call this modification multi-angle QAOA and abbreviate it ma-QAOA.

\subsection*{Convergence of ma-QAOA}
 For QAOA, the expected value of $C$ after $p$ iterations is $\langle C \rangle_p = \bra{\gamma, \beta}C\ket{\gamma, \beta}$. Let $M_p$ be the maximum of $\langle C \rangle_p$ over all angles. Then, $M_p \geq M_{p-1}$. Farhi, Goldstone, and Gutmann showed that $M_p$ tends to the maximum of the objective function, $C_\mathrm{max}$, for the CO problem being solved as $p$ tends to infinity \cite{farhi2014quantum}.
 
 We similarly define the expected value of $C$ after $p$ iterations of ma-QAOA as $\langle C \rangle_p^\mathrm{ma} = \bra{\vec{\gamma}_{\mathrm{ma}}, \vec{\beta}_{\mathrm{ma}}}C\ket{\vec{\gamma}_{\mathrm{ma}}, \vec{\beta}_{\mathrm{ma}}}$ where $\vec{\gamma}_{\mathrm{ma}} = (\vec \gamma_1, \vec \gamma_2,...\vec \gamma_p)$ and  $\vec{\beta}_{\mathrm{ma}} = (\vec \beta_1, \vec \beta_2,...\vec \beta_p)$. We also define $M_p^\mathrm{ma}$ to be the maximum of $\langle C \rangle_p^\mathrm{ma}$ over all angles. 
Clearly, $M_p^\mathrm{ma} \geq M_p$ because QAOA is the special case of ma-QAOA where $\beta_{p,u} = \beta_{p,v}$ for all $u \neq v$ and $\gamma_{p,a_i} = \gamma_{p,a_j}$ for edges $a_i \neq a_j$.

In order to show ma-QAOA gives the optimal solution to a combinatorial optimization problem, we must show $\langle C \rangle_p^\mathrm{ma}$ converges to $C_\mathrm{max}$ as $p$ tends to infinity. Convergence is the first main result of this work.

 \begin{thm}\label{thm:maconvergence}
The multi-angle quantum approximate optimization algorithm converges to the optimal solution of a combinatorial optimization problem as $p \rightarrow \infty$.
\end{thm}

The proof of convergence is given in Sec.~\ref{methods}.

\subsection*{MaxCut on star graphs}
Recall that the approximation ratio for $p$-QAOA is equal to $\langle C \rangle_p / C_\mathrm{max}$, where $C_\mathrm{max}$ is the optimal solution to the CO problem. Similarly, we define the approximation ratio for $p$ iterations of ma-QAOA as $\langle C \rangle_p^\mathrm{ma} / C_\mathrm{max}$. The approximation ratio is often used to measure the success of QAOA. 

\begin{figure}
 \centering
 \begin{tikzpicture}[scale=1.5]
\begin{scope}[every node/.style={scale=.75,circle,draw}]
    \node (A) at (0,0) {};
    \node (B) at (1,0) {};
	\node (C) at (-1,0) {}; 
	\node (D) at (0,1) {};
    \node (E) at (0,-1) {};

\end{scope}

\draw  (A) -- (B);
\draw  (A) -- (C);
\draw  (A) -- (D);
\draw  (A) -- (E);

\end{tikzpicture}
\caption{The star graph on five vertices.}
\label{fig:star}
\end{figure}
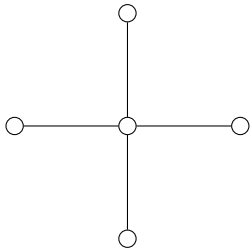

In graph theory, a star graph on $n$ vertices is a graph that consists of one vertex of degree $n-1$, called the center. All other vertices of the graph have degree one, meaning each vertex is connected to the center and only the center. An example can be seen in Fig.~\ref{fig:star}. All stars are trees, and are thus bipartite, so the optimal MaxCut solution includes all edges of the graph. In order to show ma-QAOA outperforms QAOA when solving MaxCut on star graphs, we show that $\langle C \rangle_1^\mathrm{ma} = 1$ and $\langle C \rangle_1$ tends to $0.75$ as $n$ tends to infinity. The proof is found in Sec.~\ref{methods}.

\subsection*{Computational results}

In order to test how ma-QAOA performs, we simulated the algorithm on a collection of one-hundred triangle-free 3-regular graphs with fifty vertices and one-hundred triangle-free 3-regular graphs with 100 vertices and compared the approximation ratios calculated with ma-QAOA to those of 1-QAOA. We also performed the same calculations with fifty modified $G_{n,p}$ random graphs with fifty and one-hundred vertices each. In the $G_{n,p}$ model, $n$ sets the number of nodes, and $p$ is the probability that an edge exists. In particular, we examined $G_{50, 0.08}$ and $G_{100, 0.035}$ in order to create random graphs that have average degree approximately three. After randomly generating the graphs, triangles were removed by randomly removing edges from each triangle. For these sets of triangle-free graphs we can compute $\langle C \rangle_1^\mathrm{ma}$ for large $n$ using the analytical result of Thm.~\ref{closedform}. Table~\ref{tab:averagelargegraph} shows the average approximation ratios for each collection of graphs with ma-QAOA and 1-QAOA, as well as the changes in the approximation ratio and percent change in the approximation ratio gap. This approximation ratio gap is the percent difference between one minus the approximation ratio for 1-QAOA and one minus the approximation ratio for ma-QAOA. The ma-QAOA has a higher average approximation ratio and gives a significant percent increase in approximation ratio gap for each collection of graphs. These simulations only compare 1-QAOA to ma-QAOA, however, the next set of computational results compares ma-QAOA to $p$-QAOA for $p \leq 3$ on all connected, non-isomorphic graphs.

In previous work, we determined $\langle C \rangle_1$, $\langle C \rangle_2$, and $\langle C \rangle_3$ for all connected, non-isomorphic eight vertex graphs and compiled them into an online data set \cite{lotshaw2021bfgs, lotshawdataset}. For this work, we calculated the angles that maximize $\langle C \rangle_1^\mathrm{ma}$ and compared $\langle C \rangle_p$ to $\langle C \rangle^\mathrm{ma}_1$. On average, the performance of ma-QAOA is comparable to 3-QAOA on these graphs. Table \ref{tab:average} shows that ma-QAOA has a higher average approximation ratio than 1-QAOA and 2-QAOA on all eight vertex graphs. However, the average approximation ratio for one iteration of ma-QAOA is larger than the average approximation ratio for 3-QAOA.

\begin{figure}
    \centering
    \includegraphics[scale=0.55]{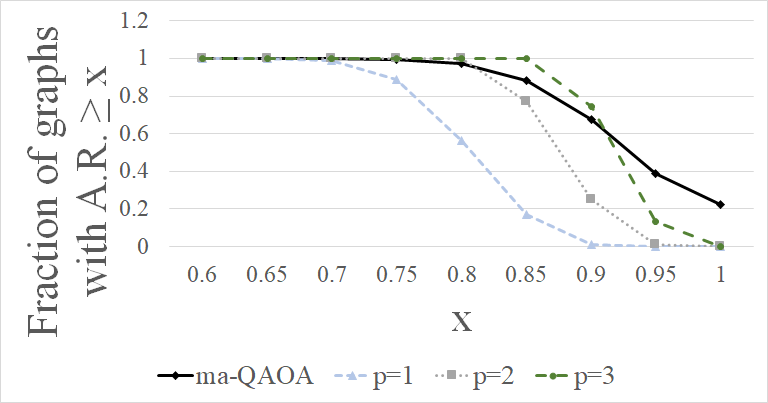}
    \caption{ Fractions of non-isomorphic, eight vertex graphs with approximation ratios (A.R.) at least $x$ for ma-QAOA and $p$-QAOA. The lines are included in order to outline the shape of each distribution.}
\label{fig:eightvertexcumulative}
\end{figure}

Figure \ref{fig:eightvertexcumulative} shows how the distribution of approximation ratios for ma-QAOA compares to the approximation ratios for up to three iterations of QAOA for all connected, non-isomorphic eight vertex graphs. The percentage of graphs with approximation ratio at least 0.95 is significantly higher with ma-QAOA than up to three levels of QAOA. The fraction of graphs with approximation ratio at least 0.85 and 0.9 is higher for 3-QAOA than ma-QAOA, however significantly more graphs have an approximation ratio of at least 0.95 with ma-QAOA.

 \begin{table*}\footnotesize
    \centering
    \begin{tabular}{|c|c|c|c|c|}
        \hline
        Graph type & Average A.R. for 1-QAOA & Average A.R. for ma-QAOA & Change in A.R. & Percent change in Gap (1-AR) \\
       
        \hline
         50 vertex 3-regular & 0.7617 & 0.8123 & 0.0506 & 21.26\% \\
        \hline
        100 vertex 3-regular &  0.7562 & 0.8000 & 0.0438 & 17.98\% \\
        \hline
        Modified $G_{50,.08}$ & 0.7554 & 0.8156 & 0.0602 & 24.65\& \\
        \hline
        Modified $G_{100,.035}$ &  0.7497 & 0.8098 & 0.0602 & 24.04\% \\
        \hline
       
    \end{tabular}
    \caption{The average approximation ratio (A.R.) for a collection of one-hundred 3-regular graphs with fifty vertices, one-hundred 3-regular graphs with 100 vertices, fifty modified $G_{50,0.08}$ random graphs, and fifty modified $G_{100,0.035}$ random graphs. 
    }
    \label{tab:averagelargegraph}
\end{table*}

 \begin{table}\footnotesize
    \centering
    \begin{tabular}{|c|c|}
        \hline
        QAOA type & Average approximation ratio for all eight vertex graphs\\
       
        \hline
         ma-QAOA & .9257\\
        \hline
        1-QAOA &  .8061 \\
        \hline
        2-QAOA &  .8767\\
        \hline
        3-QAOA & .9192\\
        \hline
       
    \end{tabular}
    \caption{The average approximation ratio for all connected, non-isomorphic graphs eight vertices. 
    }
    \label{tab:average}
\end{table}

\subsection*{Circuit Fidelity}

On fully connected hardware, the number of one-qubit unitary operators and two-qubit unitary operators per iteration of QAOA for MaxCut corresponds to the number of vertices and edges in the graph respectively. In order to solve MaxCut on all eight vertex graphs, eight one-qubit unitary operators and between seven and twenty-eight two-qubit unitary operators are needed per iteration of the algorithm, depending on the number of edges in the graph. 
Assuming a simple noise model based on Kraus-operator error channels acting after each unitary, the circuit produces a final state $\rho = F \rho_\mathrm{ideal} + (1-F)\rho_\mathrm{noise}$, where $F$ is the probability associated with the ideal noiseless evolution component $\rho_\mathrm{ideal}$  \cite{lotshaw2021resource}. The expected number of measurements to sample from the ideal distribution is $1/F$ in the worst-case, when $\mathrm{Tr}\rho_\mathrm{ideal}\rho_\mathrm{noise}=0$; more generally the number of measurements can decrease depending on the specific state and noise process, but to keep the discussion general we take the expected number of measurements as $1/F$. With this model, for $p$-QAOA or $p$-ma-QAOA, $F = (1-\epsilon_{n})^{np}(1-\epsilon_{m})^{mp}$, where
 $\epsilon_{n}$ is the expected error per single-qubit unitary, $\epsilon_{m}$ is the expected error per two-qubit unitary, $n$ is the number of qubits, and $m$ is the number of edges in the MaxCut graph. Assuming $p=1$ and an error rate of 1\% for each unitary operator, the expected number of measurements to obtain a sample from the noiseless distribution is 1.25 when $n=8$ and $m=14.4$. We choose $m=14.4$ since this is the average number of edges per graph in the collection of connected, non-isomorphic eight vertex graphs, though note each specific graph has an integer number of edges.

We find that parameter optimization with ma-QAOA yields angles of zero for a subset of the edge and vertex unitary operators. Since $\exp(-i \gamma_{p,a}C_{a})=\mathbb{I}=\exp(-i\beta_{p,v}B_v)$ when $\gamma_{p,a}=0$ and $\beta_{p,v} = 0$, all unitary operators with an angle of zero may be excluded from  the optimized circuit. The presence of fewer operators reduces the amount of noise in the circuit by decreasing the exponent of the first and second terms in $F$ by the number of vertex and edge operators that have zero angles, respectively. 
When maximizing the expected value of the cost function using ma-QAOA, a sizeable percent of angles associated with vertices and edges were zero. Table~\ref{tab:percentzero} gives the percent of zero angles, rounded to three decimal places, for each collection of graphs that were studied. These zero angles impact the fidelity of each solution, as the operators associated with them need not be implemented on hardware. Table~\ref{tab:sampleaverage} shows the ratio of the expected number of measurements needed to sample from the noiseless distribution for  $p$-QAOA relative to ma-QAOA for each collection of graphs with varying values of $\epsilon_{m}$, using the average reduction in gates for ma-QAOA from Table \ref{tab:percentzero}. Note that if the $\epsilon_m =0.05$, the number of samples increases rapidly with $p$. Since one iteration of ma-QAOA is comparable to three iterations of QAOA on eight vertex graphs, if the trend holds for larger graphs, ma-QAOA has the potential to require significantly fewer samples than QAOA. 

 \begin{table}\footnotesize
    \centering
    \begin{tabular}{|c|c|c|}
        \hline
        $n$ & Percent of $v$ with $\beta_v = 0$ & Percent of $a$ with $\gamma_a = 0$\\
       
        \hline
         8 & 15.030 & 25.449\\
        \hline
        50 (3-reg.) & 13.000 & 18.147\\
        \hline
        50 (E.R.) & 11.440 & 14.381\\
        \hline
        100 (3-reg.) & 14.690 & 19.973\\
        \hline
        100 (E.R.) & 12.900 & 16.541\\
        \hline
       
    \end{tabular}
    \caption{The percent of $\beta_v$ and $\gamma_a$, rounded to three decimal places, that are zero when optimizing ma-QAOA on the family of graphs found in the first column. 
    }
    \label{tab:percentzero}
\end{table}

 \begin{table}\footnotesize
    \centering

\begin{tabular}{|c|c|c|c|c|c|c|c|}
    \hline
    $n$ & $m$ &
    \multicolumn{3}{|c|}{$\epsilon_n = \epsilon_m = 0.01$} &
    \multicolumn{3}{c|}{$\epsilon_n = 0.01, \epsilon_m = 0.05$} \\
    & & $p=1$ & $p=2$ & $p=3$ & $p=1$& $p=2$ & $p=3$ \\
    \hline
    8 & 14.4 & 1.05& 1.32 & 1.65 & 1.22 & 2.77 & 6.28\\
    \hline
     50 (3-reg.) &75 & 1.22& 4.30 & 15.10 & 2.15 & 166.16 & $1\times 10^4$ \\
    \hline
     50 (E.R.) & 87.2 & 1.20 & 4.77 & 18.94 & 2.02 & 291.78 & $4\times 10^4$\\
    \hline
     100 (3-reg.) & 150 & 1.57 & 19.32 & 238.39 & 5.39 & $3\times 10^4$ & $2\times 10^8$ \\
    \hline
     100 (E.R.) & 167.34 & 1.50 & 22.08 & 324.26 & 4.71 & $7\times 10^4$ & $1\times 10^9$\\
    \hline
\end{tabular}
\label{tab:multicol}
    \caption{The ratio of the expected number of measurements to obtain a sample from the noiseless distribution for $p$-QAOA relative to $1$-ma-QAOA on an $n$ vertex graph with $m$ edges. }\label{tab:sampleaverage}
\end{table}

\subsection*{Computing Angles}

With a larger number of variables to optimize, the ma-QAOA method requires more classical effort to find angles that optimize the approximation ratio. However, it is not necessary to identify exact optimal angles, only to find angles that are better than QAOA angles. 

\begin{figure}
\centering
  \includegraphics[width=\linewidth,trim={0 0 1cm 0},clip]{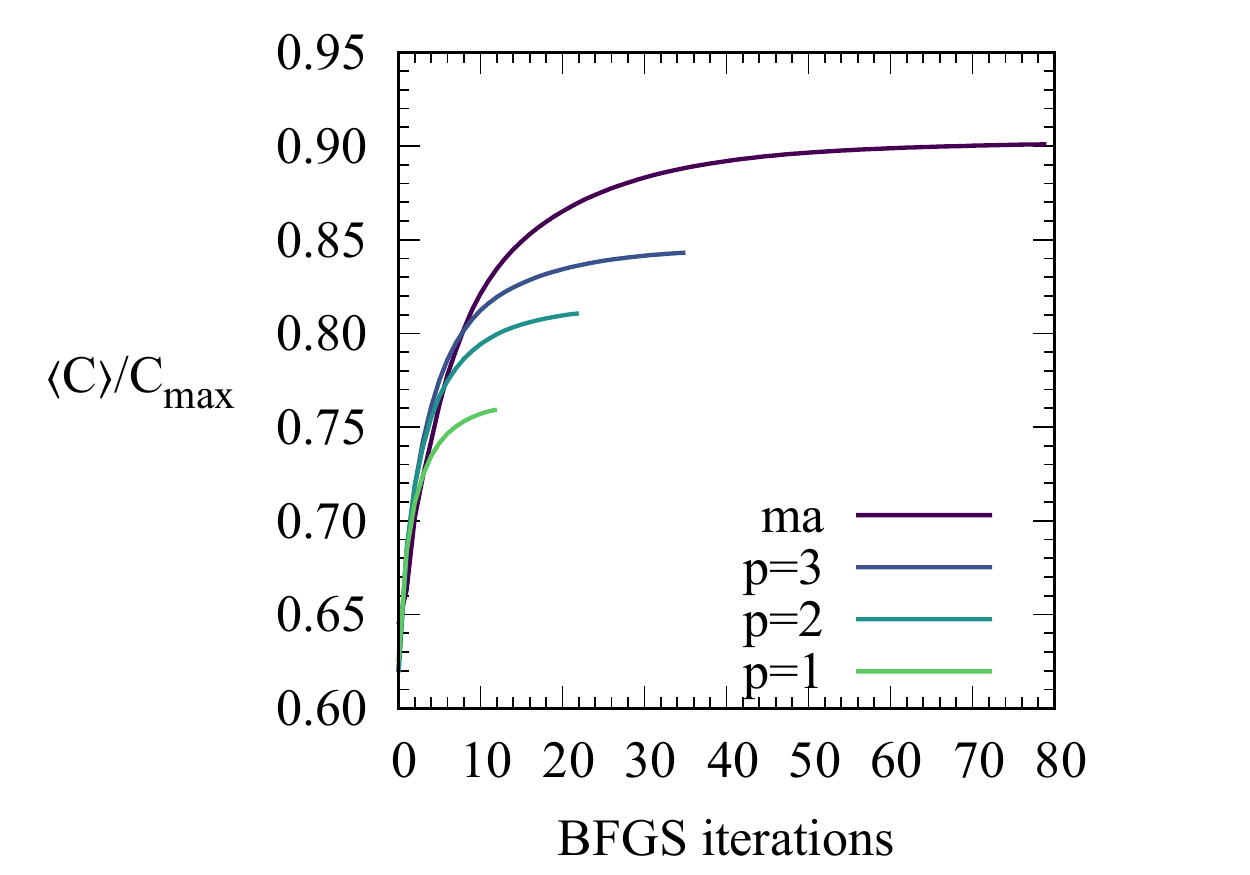}

\caption{Typical behavior of the BFGS search algorithm for ma-QAOA at $p=1$ and regular QAOA at $p=1,2,3$ on a set of 100 random eight vertex graphs.  Each curve is computed as an average over the graphs and over 100 random angle seeds for each graph.  Each curve terminates at the average final iteration of the BFGS algorithm for the dataset. }
\label{fig:BFGS_convergence}
\end{figure}

We used the BFGS algorithm to compute angles for the 8-vertex graphs; details can be found in Methods. Figure~\ref{fig:BFGS_convergence} shows how the approximation ratio improves on average across all iterations of BFGS for each ansatz studied for a random sample of eight vertex graphs. Note that after approximately ten iterations, ma-QAOA tends to achieve a higher approximation ratio than any of the $p$-QAOA. We do note that the time required to perform each iteration of BFGS is slower for ma-QAOA, as the number of gradient components is linearly dependent on the number of variables being optimized.

\subsection*{Scaling}

We assess the scalability of ma-QAOA using computed optimized $\langle C \rangle$ for sets of triangle-free Erd\H{o}s-R\'enyi and 3-regular graphs of sizes $n=50$ and $n=100$.  The computational details are given in the Methods. We compare the run times for typical graph optimizations to assess how the ma-QAOA parameter optimization time increases with graph size.

For the Erd\H{o}s-R\'enyi graphs, the time for a single optimization for $n=50$ was 0.10 seconds, for $n=100$ it was 0.46 seconds. We attribute the difference primarily to the scaling in the calculation of the gradient, which is the most expensive calculation in the optimization. Our approach computes each of the $n+m$ derivatives $\partial \langle C_{p,uv}\rangle^\mathrm{ma}/\partial \beta_{p,w}$ and $\partial \langle C_{p,uv}\rangle^\mathrm{ma}/\partial \gamma_{p,jk}$ for each of the $m$ terms $\langle C_{p,uv}\rangle^\mathrm{ma}$ in the cost function, giving a total number of terms $\sim(n+m)m$. The time to compute each term will vary with the degree of the graph, as this determines the number of cosine terms in Thm.~\ref{closedform}; however, for our graphs the degree is approximately constant hence does not affect the scaling. 
For our graphs $m \sim n$ on average, so the overall scaling is $\sim n^2$, which is consistent with the $\approx 4\times$ increase in time when $n$ is doubled from $n=50$ to $n=100$.  We attribute the remainder of the time difference to variations in the number of iterations as $n$ and $m$ increase. 

 \section{Discussion}\label{discussion}
We have shown that multi-angle QAOA converges to an optimal solution, and furthermore that $\langle C \rangle_1^\mathrm{ma} \geq \langle C \rangle_1$, as QAOA is a special case of ma-QAOA. We also show that the circuit depth for one iteration of ma-QAOA can be less than the depth for one iteration of QAOA. Additionally, the analysis of star graphs shows that there is a family of graphs that gives larger $\langle C \rangle$ for MaxCut when solved with ma-QAOA than when solved with QAOA. We find significant increases in the approximation ratio in numerical optimizations for large triangle-free graphs and over the set of all non-isomorphic graphs with eight vertices, hence fewer layers are required to reach the same performance as QAOA.  In the presence of noise, the reduction in number of layers and in the number of unitary operators per layer can significantly decrease the expected number of measurements needed to sample from the noiseless distribution.  This could be a significant advantage for computations on noisy quantum hardware.

Interestingly, some graphs do not have a significantly higher $\langle C \rangle$ when solving MaxCut with ma-QAOA versus QAOA. It would be useful to characterize for which graphs the increase in $\langle C \rangle$ from QAOA to ma-QAOA is insignificant. This would help determine the appropriate ansatz to use in order to solve MaxCut on the graph.

One drawback to ma-QAOA is that the number of classically optimized parameters is $n+m$ per layer, where $n$ is the number of vertices of $G$ and $m$ is the number of edges. From a practical standpoint, one way to solve optimal ma-QAOA angles would be to calculate $\beta$ and $\gamma$ that optimize QAOA. We can use those angles as the initial point of a BFGS search for the optimal $\beta_{p,v}$ and $\gamma_{p, a_i}$ for all vertices $v$ and edges $a_i$. Overall, however, the results seem to indicate that good paramaters can be found in polynomial time. As many combinatorial optimization problems, like MaxCut, are NP-Hard, any polynomially-bounded effort that improves performance is likely to improve performance at large scale. 

\section{Methods}\label{methods}
\subsection*{Proof of convergence}
\begin{proof}
Recall that QAOA converges to the optimal solution for a combinatorial optimization problem, which is the maximum over the objective function \cite{farhi2014quantum}.
Thus, in order to show convergence of ma-QAOA, we need only bound ma-QAOA from below by the value of QAOA. However, it is clear that the optimal expected value of the cost function for ma-QAOA can be no lower than that of QAOA, since QAOA is a special case of ma-QAOA when all $\gamma_{p,ij} = \gamma_{p,kl}$ and all $\beta_{p,a} = \beta_{p,b}$ for all edges $ij, kl$ and all vertices $a, b$.
\end{proof}
\subsection*{$\langle C \rangle$ for solving MaxCut on triangle-free graphs using ma-QAOA}
In order to prove that $\langle C \rangle_1^\mathrm{ma} = 1$ for MaxCut on star graphs, we derive a formula that calculates $\langle C \rangle_1^\mathrm{ma} $ for MaxCut on triangle-free graphs. 

 \begin{thm}\label{closedform}

Let $\beta_{p,u}' = 2\beta_{p,u}$ and $\beta_{p,v}' = 2\beta_{p,v}$ The expected value of $C$ after one iteration of ma-QAOA applied to MaxCut for triangle free graphs $G$ is 
\begin{align*}
\begin{split}
   & \bra{\vec{\gamma_1}\vec{\beta_1}} C_{uv} \ket{\vec{\gamma_1}\vec{\beta_1}} = \frac{1}{2}+ \\
   & \frac{1}{2}\sin{\gamma_{1,uv}}(\cos{\beta_{1,v}'}\sin{\beta_{1,u}'}\prod_w \cos{\gamma_{1,uw}}+ \\
   & \cos{\beta_{1,u}'}\sin{\beta_{1,v}'} \prod_{x}\cos{\gamma_{1,vx}}) 
\end{split}
\end{align*}
where $w \in Nbhd(u)\setminus v$ and $x \in Nbhd(v)\setminus u$.
\end{thm}

\noindent  The neighborhood of a vertex $x$, denoted $Nbhd(x)$, is the set of vertices $y$ such that $xy \in E(G)$. 
 \begin{proof}
 The proof of this result relies on the Pauli-solver algorithm, which is explained in detail in \cite{hadfield2018quantum}. The proof of the result is virtually identical to that for QAOA on triangle-free graphs, but we include the proof here for completeness.
 
 Consider edge $uv$ and consider acting on $C_{uv}= (1/2)(\mathbb{I}-Z_uZ_v)$ by conjugation of the mixing operator, $\prod_{i \in V}e^{-i \beta_{1,i} B_i},$ followed by conjugation of the phase operator, $\prod_{uv \in E} e^{-i \gamma_{1,uv} C_{uv}}$. 
We have that
\begin{equation}\label{conjugatetwo}
\begin{split}
& \prod_{i \in V}e^{i \beta_{1,i} B_i}Z_uZ_v\prod_{i \in V}e^{-i \beta_{1,i} B_i} =
e^{2i\beta_{1,u} X_u}e^{2i\beta_{1,v} X_v}Z_uZ_v = \\ 
&\cos{2\beta_{1,u}}\cos{2\beta_{1,v}}Z_uZ_v+ \cos{2\beta_{1,v}}\sin{2\beta_{1,u}}Y_vZ_u \\ 
& +\cos{2\beta_{1,u}}\sin{2\beta_{1,v}}Z_vY_u+\sin{2\beta_{1,u}}\sin{2\beta_{1,v}}Y_uY_v. 
\end{split}
\end{equation}
 Note that the first term commutes with $\prod_{uv \in E} e^{-i \gamma_{1,uv} C_{uv}}$, so does not contribute to the expected value. Let $V_u$ be the neighborhood of $u$ in $V(G)$. Conjugating the third term of Eqn.~\eqref{conjugatetwo} by $\prod_{uv \in E} e^{-i \gamma_{1,uv} C_{uv}}$, we get 
 
 \begin{align*}
    & \bra{s}\Upsilon^\dag Y_uZ_v\Upsilon \ket{s}= \\
     &\bra{s}e^{2i \gamma_{1,uv} C_{uv}}e^{2i \sum_{a \in V_u\setminus {v}} \gamma_{1,ua} C_{ua} }Y_uZ_v\ket{s} = \\
    &\bra{s}e^{-i \gamma_{1,uv} Z_uZ_v}e^{-i \sum_{a \in V_u \setminus {v}} \gamma_{1,ua} Z_uZ_a }Y_uZ_v\ket{s} = \\
    &\bra{s}(\mathbb{I}\cos{\gamma_{1,uv}}-i\sin{\gamma_{1,uv}})Z_uZ_v \\
    & \prod_{a \in V_u}(\mathbb{I}\cos{\gamma_{1,ua}}-i\sin{\gamma_{1,ua}}Z_uZ_a)Y_uZ_v\ket{s} = \\
    &-\sin{\gamma_{1,uv}}\prod_{a \in V_u}\cos{\gamma_{1,ua}}, 
 \end{align*}

 \noindent where $\Upsilon = e^{-i \gamma_{1,uv} C_{uv}}e^{-i \sum_{a \in V_u\setminus {v}} \gamma_{1,ua} C_{ua}}$, and $\Upsilon^\dag$ is its Hermitian conjugate. By symmetry, the term for $Z_uY_v$ is $-\sin{\gamma_{1,uv}}\prod_{b \in V_v\setminus{u}}\cos{\gamma_{1,vb}}$, where $V_v$ is the neighborhood of $v$ in $V$. Factoring in the coefficient $-1/2$ of $Z_uZ_v$ in $C_{uv}$ gives the final two terms in the theorem.

 Now, let us conjugate the last term of Eqn.~\eqref{conjugatetwo}. Doing so, we get
 
 \begin{align*}
     & \bra{s}e^{i \sum_{gh \in E} \gamma_{1,gh} C_{gh}}Y_uY_v e^{-i \sum_{gh \in E} \gamma_{1,gh} C_{gh}} \ket{s} =\\
    & \bra{s}\prod_{a \in V_u\setminus v}(\mathbb{I}\cos{\gamma_{1,ua}}- i\sin{\gamma_{1,ua}}Z_uZ_a)\times\\
    & \prod_{b \in V_v\setminus u}(\mathbb{I}\cos{\gamma_{1,vb}}-i\sin{\gamma_{1,vb}}Z_vZ_b)Y_uY_v\ket{s}  \\
 \end{align*}
 
 The simplest terms that contribute to the expected value are of the form 
 \begin{align*}
     \sin{\gamma_{1,uc}}\sin{\gamma_{1,vc}}\prod_{x \neq y}\cos{\gamma_{1,ux}}\cos{\gamma_{1,vy}}
 \end{align*}
 
\noindent and there are $f$ of these where $f$ is the number of triangles containing $uv$. The higher order terms only contribute to the expected value if there are triangles in the graph. Thus, the last term of Eqn.~\eqref{conjugatetwo} contributes nothing to the expected value of triangle-free graphs.
 
 Combining these expressions gives the theorem.
 \end{proof}
 
 \subsection*{Star graphs}
 
 First, we will show that $\langle C_{ij} \rangle$ approaches $0.75$ as $n$ tends to infinity for QAOA. Since there are $n-1$ edges in a star on $n$ vertices, this implies $\langle C \rangle$ tends to $0.75(n-1)$. Additionally, $n-1$ is the size of the optimal MaxCut solution, so $\langle C \rangle_1/C_\mathrm{max} = 0.75$.
 
  Wang, Hadfield, Jiang, and Rieffel showed that \cite{wang2018quantum}
 
 \begin{equation}\label{evformula}
\begin{split}
     &\langle C_{ij} \rangle_1 = \frac{1}{2}+\frac{1}{4}(\sin{4\beta}\sin{\gamma})(\cos^{d}{\gamma}+\cos^{e}{\gamma}) \\
     & -\frac{1}{4}(\sin^2{2\beta}\cos^{d+e-f}{\gamma})(1-\cos^{f}{2\gamma})
\end{split}
\end{equation}
\noindent  where $d$ is the $\deg(i)-1$, $e$ is the $\deg(j)-1$ and $f$ is the number of triangles containing edge $ij$ \cite{wang2018quantum, hadfield2018quantum}.

Let us consider the above formula applied to a star graph. Without loss of generality, let $j$ be the center of each star. Then $d = 0$, $e = n-2$, and $f = 0$, since star graphs are trees. For each edge of the star, Eq.~\eqref{evformula} reduces to
 \begin{equation*}
     \langle C_{ij} \rangle_1 = \frac{1}{2}+\frac{1}{4}(\sin{4\beta}\sin{\gamma})(1+\cos^{n-2}{\gamma}).
 \end{equation*}
We set $\beta = \pi/8$, which implies $\sin{4\beta} = 1$, since only one trigonometric function has $\beta$ as an argument. As $n$ tends to infinity, note $\cos^{n-2}{\gamma}$ tends to zero unless $\gamma = k\pi$ for some $k \in \mathbf{N}$. However, if $\gamma = k\pi$, $\sin{\gamma} =0$. Thus, this quantity is maximized when $\gamma \neq k\pi$, which implies $\langle C_{ij} \rangle_1$ tends to $0.75$ for star graphs.

In order to prove  $\langle C \rangle^\mathrm{ma} = n-1$ for ma-QAOA on star graphs, we examine Thm.~\ref{closedform}. Without loss of generality, let $u$ be a leaf vertex and $v$ be the center. Note that the first product is empty, since the leaf vertices have no neighbors except the center. Thus, Thm.~\ref{closedform} reduces to

\begin{align*}
   & \bra{\vec{\gamma_1}\vec{\beta_1}} C_{uv} \ket{\vec{\gamma_1}\vec{\beta_1}} = \frac{1}{2}+ \\
   & \frac{1}{2}\sin{\gamma_{1,uv}}(\cos{\beta_{1,v}'}\sin{\beta_{1,u}'}+\cos{\beta_{1,u}'}\sin{\beta_{1,v}'} \prod_{x}\cos{\gamma_{1,vx}})
\end{align*}

 \noindent Now, recall $\bra{\vec{\gamma_1}\vec{\beta_1}} C_{uv} \ket{\vec{\gamma_1}\vec{\beta_1}} \leq 1$, as two vertices that have an edge between them add one to the objective function if they are in different sets. In order to obtain equality, we can set $\gamma_{1,uv} = \pi/2$, as it is an argument for only a single sine term. Next, note that if either term in the parenthesis is one, the other must be zero. Also, setting one term equal to one allows gives an expected value of one for the edge.  Let $\beta_{1,u}' = \pi/2$ and $\beta_{1,v}' = 0$. Then $\cos{\beta_{1,v}'} = \sin{\beta_{1,u}'}= 1$ while  $\cos{\beta_{1,u}'} = \sin{\beta_{1,v}'}= 0$. Thus, the first term in the parenthesis is one and the second is zero. This allows us to set $\gamma_{1,vx} = \pi/2$ for all $x \in Nbhd(v)$. Since each of the $n-1$ edges in the star are described similarly, $\langle C \rangle_1^\mathrm{ma} = n-1$ for all $n$. The size of the optimal cut on a star graph is $n-1$, so $\langle C \rangle_1^\mathrm{ma}/C_\mathrm{max} = 1$.
 
 \subsection*{Setup for Computational Results}
 
 In order to calculate the angles that maximize $\langle C \rangle_p$ and $\langle C\rangle_1^\mathrm{ma}$ for the eight vertex graphs, we used the Broyden-Fletcher-Goldfarb-Shanno (BFGS) algorithm \cite{NumericalRecipesBFGS}. The algorithm inputs an initial collection of angles and then uses a numerical gradient and second order approximate Hessian to find angles that converge to local maxima of $\langle C \rangle_p$ and $\langle C\rangle_1^\mathrm{ma}$. For the eight vertex graphs, 100 random seeds were used to optimize $\langle C \rangle^\mathrm{ma}_1$.  The results for the $\langle C \rangle_p$ were taken from the online dataset \cite{lotshawdataset} of Ref.~\cite{lotshaw2021bfgs}, where we performed an exhaustive analysis of QAOA performance on small graphs.  These used 50 seeds for $p=1$, 100 seeds for $p=2$, and 1,000 seeds for $p=3$.

 For the 50 and 100 node graphs, we used the method of moving asymptotes (MMA) algorithm \cite{nlopt,nloptmma}, but note that calculations with BFGS gave similar results.  The $\langle C\rangle_1$ were computed using Eq.~(\ref{evformula}) and the reported results were taken as the best from 1000 initial seeds in MMA optimizations.  The $\langle C\rangle_1^\mathrm{ma}$ were computed with Theorem \ref{closedform} and MMA optimization.  We report results as the best optimized values from 1,000 seeds at $n=50$ and from 100 seeds at $n=100$.


\section*{Acknowledgements}
This work was supported by DARPA ONISQ program under award W911NF-20-2-0051. J. Ostrowski acknowledges the Air Force Office of Scientific Research award, AF-FA9550-19-1-0147. G. Siopsis acknowledges the Army Research Office award W911NF-19-1-0397. J. Ostrowski and G. Siopsis acknowledge the National Science Foundation award OMA-1937008.

\bibliographystyle{unsrt}
\bibliography{references}

\end{document}